\documentclass[prb,preprint]{revtex4-1}

\usepackage{amsmath}  
\usepackage{amsfonts} 
\usepackage[colorlinks,urlcolor=blue,linkcolor=blue,citecolor=blue]{hyperref}
\usepackage{graphicx} 
\usepackage{subfigure}
\begin{document}
\title{On the nonlinearity of a tuning fork}
\author{Lintao Xiao}
\affiliation{School of Physics, Nanjing University, Nanjing, P.R. China, 210093}

\author{Chenyu Bao}
\affiliation{School of Physics, Nanjing University, Nanjing, P.R. China, 210093}

\author{Qiuhan Jia}
\affiliation{School of Physics, Nanjing University, Nanjing, P.R. China, 210093}

\author{Haoyang Wu}
\affiliation{School of Physics, Nanjing University, Nanjing, P.R. China, 210093}

\author{Huijun Zhou}
\affiliation{School of Physics, Nanjing University, Nanjing, P.R. China, 210093}

\author{Sihui Wang}\email{wangsihui@nju.edu.cn}
\affiliation{School of Physics, Nanjing University, Nanjing, P.R. China, 210093}
\date{\today}
\begin{abstract}
    Tuning fork experiments at the undergraduate level usually only demonstrate a tuning fork’s linear resonance. In this paper, we introduce an experiment that can be used to measure the nonlinear tuning curve of a regular tuning fork. Using double-grating Doppler interferometry, we achieve measurement accuracy within ten microns. With this experiment setup, we observe typical nonlinear behaviors of the tuning fork such as the softening tuning curve and jump phenomena. Our experiment is inexpensive and easy to operate. It provides an integrated experiment for intermediate-level students and a basis for senior research projects.
\end{abstract}

\maketitle
\section{Introduction}

Tuning forks are frequently used as stable frequency references for musical and timing purposes. Their mechanical and acoustical properties can be found in many papers about linear resonance,\cite{Miller1950} vibration modes\cite{Rossing1992} and sound field radiation.\cite{Russell2000} Nowadays, the quartz tuning forks (QTF) have become the core components of various sensors and microscopies\cite{Friedt2007} such as SNOM (scanning near-field optical microscopy),\cite{Gao2014} AFM (atomic force microscope)\cite{Li2016} and other microelectromechanical systems (MEMS).\cite{Lee2008,Sarrafan2018,Wang2018,Hafiz2018} Both regular and micro tuning forks exhibit nonlinear behaviors. Thomas Rossing et al. found the non-linear relationship between the amplitudes of the second and other small harmonics and the fundamental amplitude when measuring the free vibration of their tuning fork\cite{Rossing1992}. Xuefeng Wang et al. studied the resonant frequency shift of a T-shaped tuning fork micro-resonator which shows softened nonlinearity.\cite{Wang2018} In most cases in existing literature, experiment phenomena and conditions are not readily connected to appropriate mathematical models; thus many aspects of their nonlinearity remain unclear. 

In this paper, we use laser Doppler interferometry to measure the tuning fork’s principle resonance tuning curve. We find typical nonlinear behaviors such as the softening tuning curve and jump phenomena. In Sec. II, we discuss the nonlinear behaviors of the tuning forks and apply an oscillation model containing both quadratic and cubic terms in the restoring force. In Sec. III, we introduce the double-grating Doppler interferometry experiment that achieves measurement accuracy within ten microns. In Sec. IV, we observe the softening resonance curve and jump phenomena in the principle mode of a tuning fork. Our experiment studies the nonlinear dynamics of a tuning fork using laser Doppler interferometry, providing an integrated experiment for intermediate-level students. It also serves as a basis for senior research projects to further explore the phenomena.

\section{Modeling the nonlinearity}
The current model of tuning forks that approximates the tines as cantilever or free-free beams provides satisfactory explanation for the vibration modes and their corresponding frequencies. For in-plane symmetric modes, the modal frequencies are,\cite{Rossing1992}
\begin{align}\label{eq:1}
f_n=\left(\frac{\pi a}{16L^2}\right)\sqrt{\frac{E}{3\rho}}[{1.194}^2,{2.988}^2,5^2,7^2,\ldots,{(2n-1)}^2,],
\end{align}
where $a$ is the thickness of the fork’s beam, $L$ is the length of the tines, $E$ is the Young elastic modulus and $\rho$ is the density of the fork’s material. Note that the frequency of the second symmetric mode predicted by Eq. (\ref{eq:1}) is over 6 times ($\frac{{2.988}^2}{{1.194}^2}$ times) the principal mode $f_0$, much higher than small harmonic frequencies $2f_0$, $3f_0$, etc. that exist in a tuning fork’s spectrum. Apparently, these harmonics should be attributed to nonlinearity.

\begin{figure}[t]
\begin{center}
\subfigure[]{\label{fig:1a}
\includegraphics[width=0.4\textwidth]{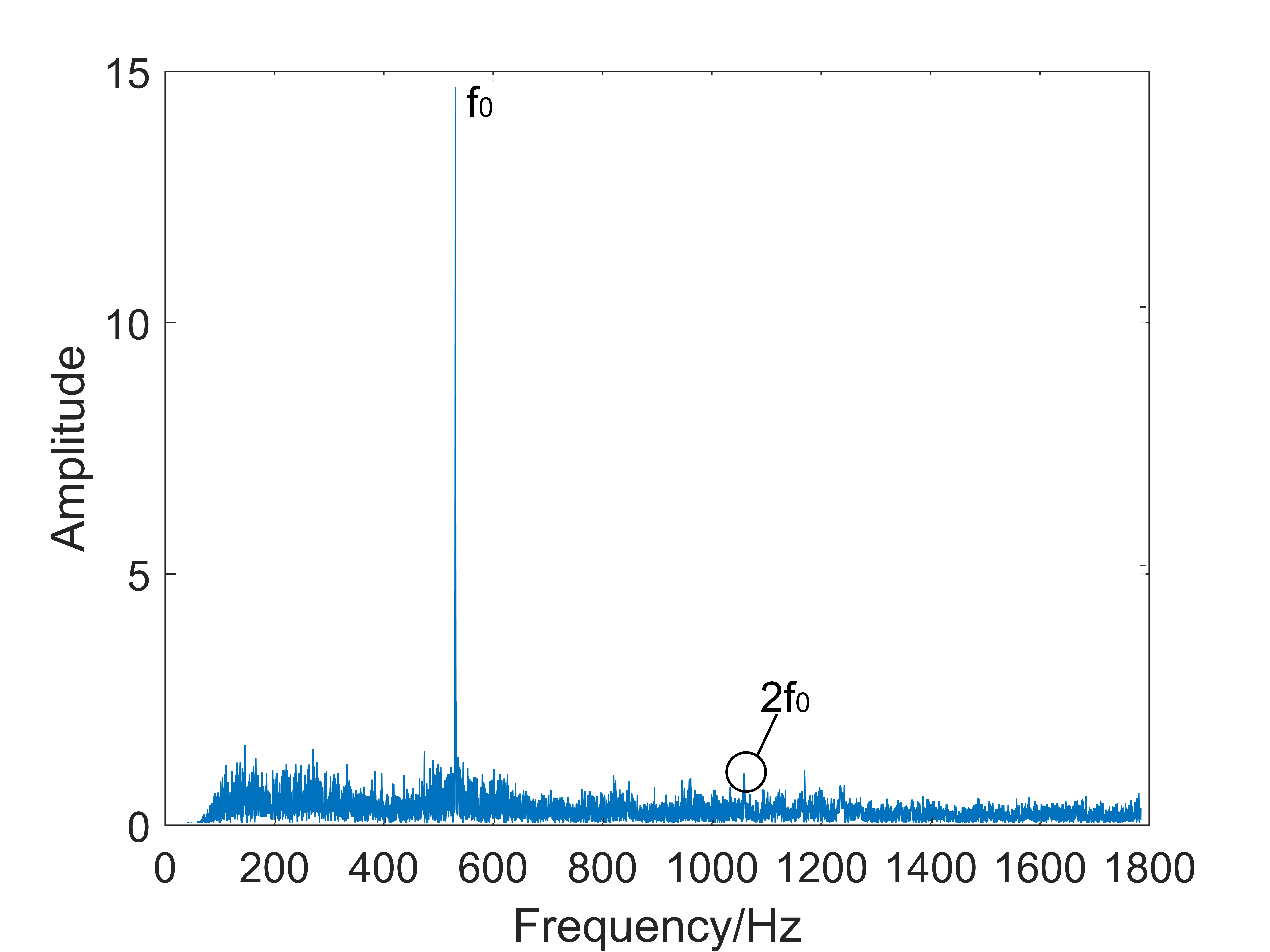}
}
\subfigure[]{\label{fig:1b}
\includegraphics[width=0.4\textwidth]{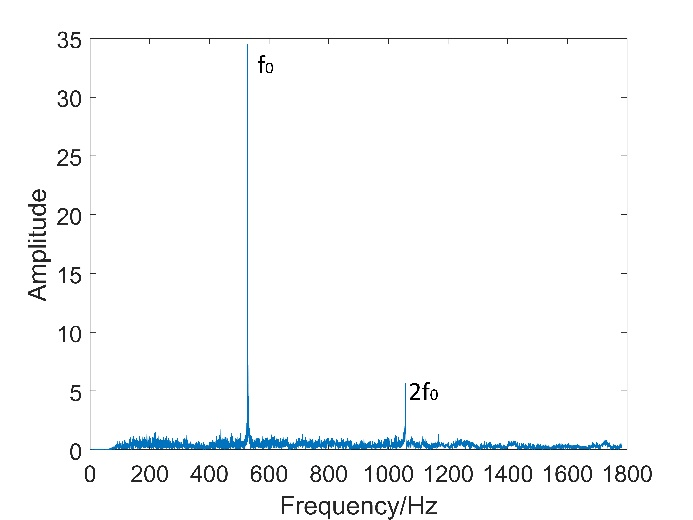}
}
\caption{ Sound spectrums radiated by a 528 Hz tuning fork for (a) a normal blow; (b) a hard blow. }
\label{fig:1}
\end{center}
\end{figure}

In fact, the motion of each tine of a fork resembles that of a cantilever beam only in linear regime. A vibrating cantilever beam is symmetric and usually modeled by the Duffing equation which contains a cubic restoring force\cite{cite10}. In a cantilever beam model, the beam’s principal mode shows hardening spring behaviors, and its vibration spectrum has a prominent third harmonic component\cite{Wagner1965}. However, these predictions based on a cantilever beam model are contradictory to experimental observations. In Section IV, our experiment shows that the principal mode exhibits softening rather than hardening spring behaviors. We also measured the free vibration spectrums radiated by a 528 Hz steel tuning fork, see Fig. \ref{fig:1}. We recorded the sound with the microphone on a pair of EarPods and processed the audio signal with Matlab. We were unable to use the same tuning fork described in Section IV, because this part of experiment was done at home due to the COVID-19 pandemic. The second harmonic component can be seen with a normal blow and becomes obvious with a hard blow on the tuning fork. However, the third harmonic component is very weak even under a hard blow.

The tines of a real-world tuning fork are geometrically and physically asymmetric. To describe the asymmetry of the restoring force, we apply a model containing both quadratic and cubic terms. The equation for free vibration is, 
\begin{align}\label{eq:2}
\ddot{y}+{\omega_0}^2y=-\alpha y^2-\beta y^3,
\end{align}
where $\omega_0$ is the natural angular frequency under linear approximation, $y$ is an effective displacement for the equivalent oscillator of a vibration mode, and $\alpha$ and $\beta$ are nonlinear quadratic and cubic stiffness parameters. This nonlinear model was solved in Landau’s textbook\cite{cite12}. For the free vibration, the amplitude for second harmonic component is $\frac{\alpha b^2}{{6\omega_0}^2}$, i.e. the amplitude is proportional to the square of fundamental amplitude $b$. In experiment, the amplitude of the strong second harmonic component is nearly proportional to the square of fundamental amplitude\cite{Rossing1992}, which clearly indicates the existence of such quadratic nonlinearity.

Now consider forced vibrations; the oscillator is governed by the following equation:  
\begin{align}\label{eq:3}
\ddot{y}+2\lambda_d\dot{y}+{\omega_0}^2y=-\alpha y^2-\beta y^3+F\cos\mathrm{\Omega t}, 
\end{align}
where $F$ and $\Omega$ are the amplitude and angular frequency of the harmonic driving force respectively, and $\lambda_d$ is the damping coefficient. In our experiment, $y$ is the displacement of the point on the fork tine where the grating is attached.

The amplitude frequency response for Eq. (\ref{eq:3}) is given by 
\begin{align}\label{eq:4}
b^2\left[\left(\varepsilon-\kappa b^2\right)+{{\lambda_d}^2}^2\right]=\frac{F^2}{4{\omega_0}^2} ,  
\end{align}
where $b$ is the amplitude of the forced vibration, $\varepsilon=\mathrm{\Omega}-\omega_0$, and  $\kappa=\frac{3\beta}{8\omega_0}-\frac{3\alpha^2}{8{\omega_0}^3}$. 

\begin{figure}[t]
\begin{center}
\subfigure[]{\label{fig:2a}
\includegraphics[width=0.4\textwidth]{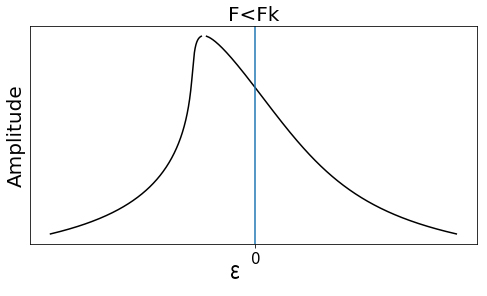}
}
\subfigure[]{\label{fig:2b}
\includegraphics[width=0.4\textwidth]{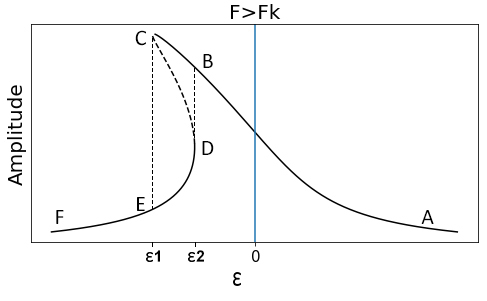}
}
\caption{Typical frequency response curves for a soften spring when (a) $F<F_k$; (b) $F>F_k$, where $F_k$ is the critical driving amplitude for a bistable state to exist.}
\label{fig:2}
\end{center}
\end{figure}

  The resonance frequency shift derived from this model is  
\begin{align}\label{eq:5}
\varepsilon_{resonance}=\kappa b^2. 
\end{align}
We see that a positive $\kappa$ corresponds to a hardening spring behavior, a negative $\kappa$ corresponds to a softening spring behavior. Typical frequency response curves for $\kappa<0$ are plotted in Fig. \ref{fig:2} according to Eq. (\ref{eq:4}). From $\kappa=\frac{3\beta}{8\omega_0}-\frac{3\alpha^2}{8{\omega_0}^3}$,  we know that the quadratic term always contributes to the softening effect regardless the sign of $\alpha$. A tuning fork shows softening behavior when the quadratic term dominates, even though the cubic nonlinearity is a hardening one.

Fig. 2(a) shows a typical frequency response curve for a soften spring when $F<F_k$, where ${F_k}^2=\frac{32{\omega_0}^2{\lambda_d}^3}{3\sqrt3\left|\kappa\right|}$. $F_k$ is the critical driving amplitude for a bistable state to appear. As shown in Fig. \ref{fig:2b}, when driving amplitude $F>F_k$, the amplitude frequency response curve becomes multivalued. For each value of frequency $\varepsilon$ in the interval between the frequencies $\varepsilon_1$ and $\varepsilon_2$, there are three steady-state solutions. Among them the middle one on CD (dotted line) is unstable, while the other two (BC and DE) are stable. Hence, as the driving frequency decreases quasi-statically from point A, the amplitude increases along the curve ABC until frequency $\varepsilon_1$ is reached. A slight frequency decrease at frequency $\varepsilon_1$ causes a spontaneous jump from C down to E. As the frequency continues to decrease beyond $\varepsilon_1$, the amplitude decreases along the curve EF. When the driving frequency increases quasi-statically from point F, the amplitude increases along the curve FED until frequency $\varepsilon_2$ is reached. A slight frequency increase at frequency $\varepsilon_2$ causes a spontaneous jump from D up to B. As the frequency increases further beyond $\varepsilon_2$, the amplitude decreases along the curve BA. In Section IV, we will verify these resonance properties and determine the parameters $\kappa$ and $\lambda_d$ experimentally.
\section{Experimental method}
\begin{figure}[t]
\begin{center}
\includegraphics[width=0.8\textwidth]{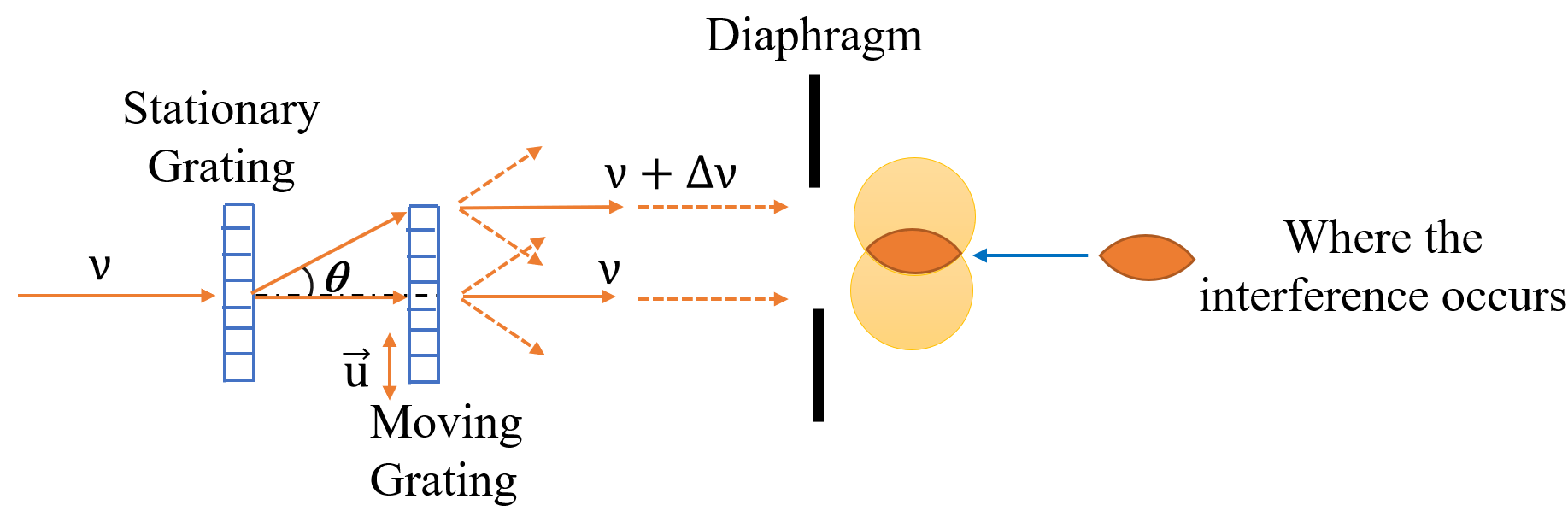}
\caption{Double-grating Doppler interferometry.}
\label{fig:3}
\end{center}
\end{figure}

In our experiment, we use double-grating Doppler interferometry\cite{Zhenhui2004,Zhang2009,Chen2016} to measure the vibration of a tuning fork, see Fig. \ref{fig:3}. Two identical diffraction phase gratings are placed parallel to each other. One is stationary, and the other is attached to one of the fork tines. 

When a beam of light is incident normally to the stationary grating plane, the light is diffracted and the $n$-th order diffracted angle $\theta_n$ satisfies $d\sin\theta_n=n\lambda$, where $\lambda$ is the wavelength of the laser and d is the grating constant. Then the beams of the diffracted light illuminating the moving grating will be diffracted again and their frequencies will shift due to the laser Doppler effect. The frequency shift is given by\cite{Bahrmann1977}
\begin{align}\label{eq:6}
\frac{\Delta \nu}{\nu} = \frac{u}{c}(\sin\theta+\sin\varphi)
\end{align}
where $\nu$ is the frequency of the light and $c$ is the speed of light; $\theta$ is the angle of incidence to the moving grating, which equals the diffraction angle of the stationary grating; $\varphi$ is the angle of diffraction of the moving grating; and $u$ is the velocity of the moving grating that is parallel to the grating plane.

In the experiment, the photodetector is placed in alignment with the incident light such that only light with angle $\varphi = 0$ can be received. For a grating with diffraction angle $\varphi=0$ and incidence angle $\theta$, the $n$-th diffraction maxima also satisfies $d\sin\theta_n=n\lambda$. Because the two gratings have the same grating constant and the same $\theta_n$, they have the same diffraction order $n$. We refer to the beam with diffraction order $n$ as beam 1. The frequency shift of beam 1 is $\Delta\nu_n=\frac{\nu u\sin\theta_n}{c}=\frac{u}{d}n$. Similarly, the frequency shift of an adjacent order (“beam 2”) is $\frac{u}{d}(n+1)$. Since the spacing between the two gratings is small, the overlapping facula of the adjacent orders interferes. The interference of beam 1 and beam 2 with different frequencies produces an optical beat whose frequency is $\nu_b=\frac{u(t)}{d}$. 

Since the beat frequency as a function of time is proportional to the velocity of the vibrating grating, the phase of the optical beat is $\varphi\left(t\right)=\int_{0}^{t}{2\pi\nu_b\left(\tau\right)d\tau}+\varphi_o=\frac{2\pi y\left(t\right)}{d}+\varphi_o$. Suppose that the motion of the grating is harmonic with angular frequency $\omega$ and amplitude b, i.e., $y\left(t\right)=b\cos(\omega t)$, then the light intensity received by the photodetector is
\begin{align}\label{eq:7}
I\left(t\right){=I}_o\cos{\left(\frac{2\pi y\left(t\right)}{d}+\varphi_o\right)}=I_o\cos{\left(\frac{2\pi b\cos(\omega t)}{d}+\varphi_o\right)},  
\end{align}
where $I_0$ is the intensity of the optical beat, and $\varphi_o$ is the initial phase.

\begin{figure}[!t]
\begin{center}
\subfigure[]{\label{fig:4a}
\includegraphics[width=0.3\textwidth]{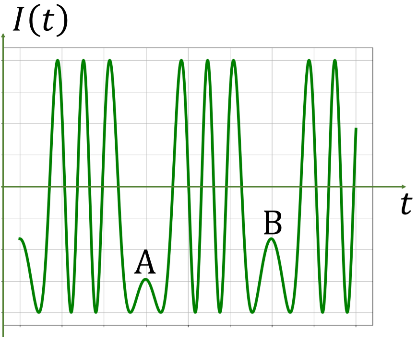}
}
\subfigure[]{\label{fig:4b}
\includegraphics[width=0.3\textwidth]{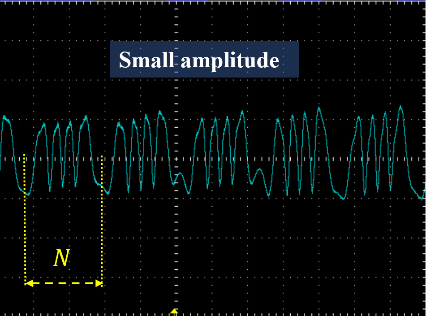}
}
\subfigure[]{\label{fig:4c}
\includegraphics[width=0.327\textwidth]{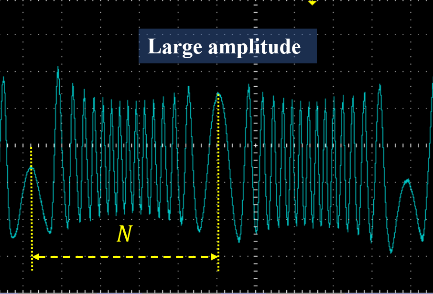}
}
\caption{ Images of harmonic motion detected by double-grating Doppler interferometry. (a) Theoretical image for $\frac{2\pi b}{d}=3.3\pi$, $\varphi_o=0.2$. (b), (c) Experimental images with small (b) and large (c) amplitudes.}
\label{fig:4}
\end{center}
\end{figure} 

A theoretical image of a typical harmonic motion detected by double-grating Doppler interferometry is shown in Fig. \ref{fig:4a}. The curve is periodic: the half period from A to B, where $\frac{d\varphi}{dt}=0$, corresponds to half period of the vibration from $t_k\equiv\frac{k\pi}{\omega}$ to $t_{k+1}\equiv\frac{(k+1)\pi}{\omega}$. The number of extrema $N$ within each half period corresponds to phase change of $\pi N\approx\varphi\left(t_{k+1}\right)-\varphi\left(t_k\right)=4\pi\frac{b}{d}$. Thus the vibration amplitude can be measured by counting the extrema 
\begin{align}\label{eq:8}
b\approx\frac{Nd}{4}.
\end{align}
The measurement accuracy is about $d/4$, reaching the range of 10 micron in our experiment. Figs.\ref{fig:4b} and \ref{fig:4c} are typical experimental images with different amplitudes. The relative uncertainty is smaller for larger amplitudes. 

\section{Experimental results }
In the experiment, we use a 512-Hz steel tuning fork. The grating constants for the gratings are $d=5.0\times 10^{-5}$ m. One grating is attached to the end of one tine of the tuning fork to measure its transverse vibration. We put an electromagnetic coil near the middle of the same tine to drive the steel tuning fork with a sinusoidal current with angular frequency $\Omega$. The sinusoidal current was biased to avoid frequency doubling of the driving force. For simplicity, we approximate the driving force as $F\cos\Omega t$. 

\begin{figure}[!t]
\begin{center}
\subfigure[]{\label{fig:5a}
\includegraphics[width=0.6\textwidth]{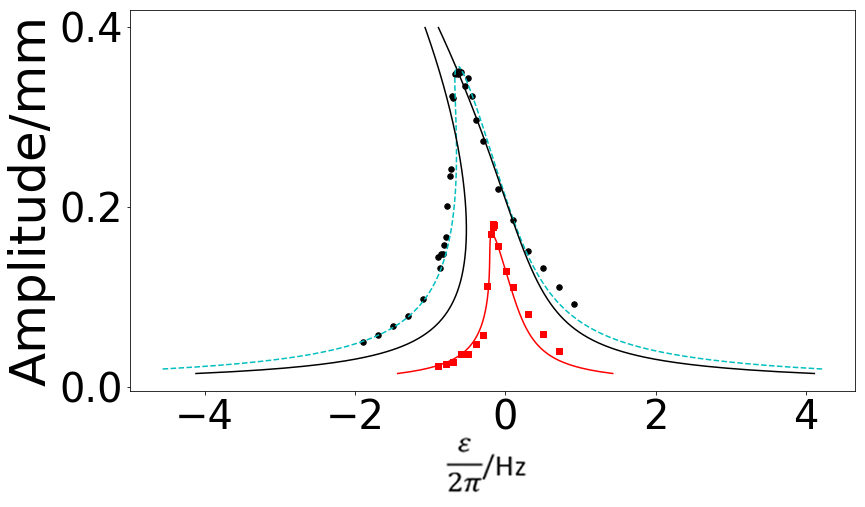}
}
\subfigure[]{\label{fig:5b}
\includegraphics[width=0.75\textwidth]{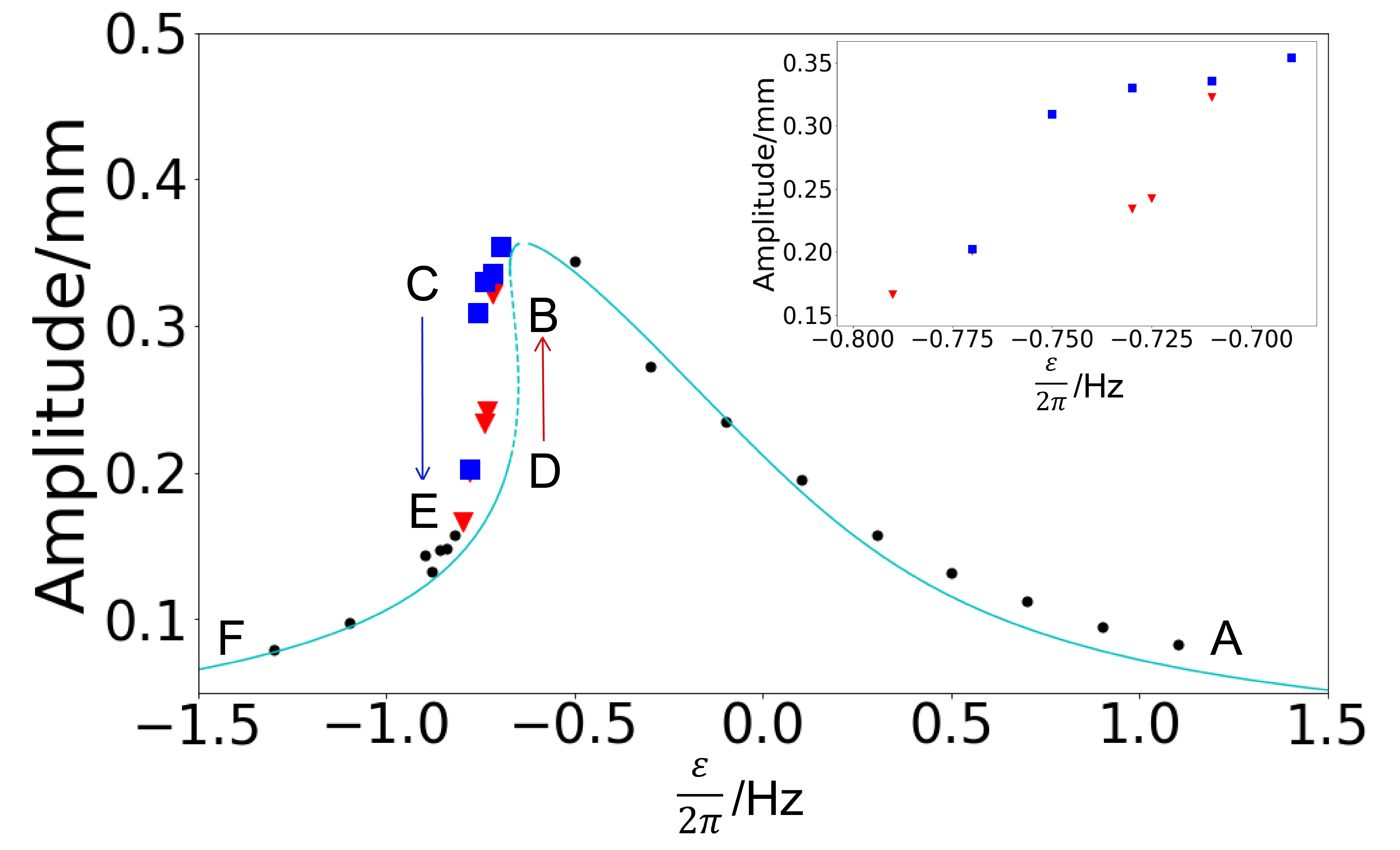}
}
\caption{Amplitude frequency response curves (a) The data plotted in red squares and black dots are results at driven amplitude $F_1$  and $F_2=25F_1$ respectively. The solid curves are theoretical results with $\kappa=-6.14\textrm{mm}^{-2} \textrm{s}^{-1}$ and $\lambda=0.25\textrm{s}^{-1}$. The broken curve is for 25$F_1$ with $\kappa=-3.66\textrm{mm}^{-2}\textrm{s}^{-1}$and $\lambda=0.49\textrm{s}^{-1}$.   (b) The “jump phenomena” at $F_2$. The “jump up” occurs above -0.73Hz (510.16Hz), “jump down” occurs below -0.75Hz (510.14Hz).}
\label{fig:5}
\end{center}
\end{figure}

Fig. \ref{fig:5a} and \ref{fig:5b} are the amplitude frequency response curves of the tuning fork, where $\epsilon=\Omega-\omega_0$. The natural angular frequency we measured from the sound spectrums is $\omega_0=2\pi\times 510.89$Hz which is lower than the labeled value after the grating is attached. In Fig. \ref{fig:5a}, the data plotted in red squares and black dots are results at driven amplitudes $F_1$ and 25$F_1$  respectively, where $F_1$ is a standard driven amplitude when the driving current amplitude $I_A=20$mA. The solid curves are theoretical results given by fitting the data with Eq. (\ref{eq:4}) with coefficients $\kappa=-6.14\textrm{mm}^{-2} \textrm{s}^{-1}$ and $\lambda=0.25\textrm{s}^{-1}$.

At driving amplitude $F_1$ the theoretical curve fits the data points well, but at 25$F_1$, the solid theoretical curve deviates from the data points systematically. In the figure, the broken curve fitted for $25F_1$ with $\kappa=-3.66\textrm{mm}^{-2} \textrm{s}^{-1}$ and $\lambda=0.49\textrm{s}^{-1}$ fits the data much better. The variation of coefficients might be caused by the driving force applied to the tines which was simplified as a linear force.

The resonance curves show that obvious softening spring behaviors and jump phenomena are observed when the driving amplitude exceeds 25$F_1$. The detail near the resonant peak is plotted in Fig. \ref{fig:5b} for driving amplitude $25F_1$. The data points in the inset correspond to the bi-stable region near BCED in Fig. \ref{fig:2b}. In this region, the data points as the driving frequency increases and decreases are indicated with red triangles and blue squares, respectively, and the curve is the theoretical result fitted for $\kappa=-3.66\textrm{mm}^{-2} \textrm{s}^{-1}$ and $\lambda=0.49\textrm{s}^{-1}$. When the driving frequency increases from point F, the amplitude increases along the curve and experiences a jump from D up to B above frequency -0.73Hz ($\Omega/2\pi =510.16$Hz). When the frequency decreases from point A, the amplitude increases along the curve and experiences a downward jump from C to E below frequency -0.75Hz ($\Omega/2\pi  =510.14$Hz). The amplitude response curve forms a hysteresis loop BCED as indicated by the arrows in Fig. \ref{fig:5b}. Even though the frequency region for the bi-stability is narrow, the amplitude jump phenomenon is impressive to the observer indeed. The jump up and jump down are accompanied with a sudden change in the loudness and a perturbation on oscilloscope signals.

It is difficult to determine the coefficients $\alpha$, $\beta$ or $\kappa$ and $\lambda_d$ from the physical parameters in our mathematical model. Nevertheless, we can still investigate resonance properties and determine these parameters experimentally. The fact that the second harmonic component is much stronger than the third harmonic component indicates that the quadratic term prevails in our tuning fork. Quantitative results may be obtained through more detailed measurement. However, in a forced vibration system like this, nonlinearity has various origins, such as geometrical, mechanical or electromagnetic nonlinearity. The parameters are sensitive to minor altering in the symmetry physically or geometrically. For example, bending the tines inward inhibits the second harmonic\cite{Rossing1992}. Driving or detecting components such as coils or electrodes coupled to the tines might also introduce additional nonlinearity to the restoring force, the driving force as well as to the damping. In these cases, the signs and magnitudes of coefficients might be altered, and the behaviors of the oscillator might change as well. In our experiment, the nonlinearity parameter and damping coefficient change as the driving force increases. By adjusting the driving or detecting parameters such as the amplitude or the bias of the driving current appropriately, the nonlinear coefficients could be controlled according to practical needs.
\section{Conclusion}
In conclusion, we investigated the nonlinear behaviors of a regular tuning fork. In theory, we apply an oscillation model containing both quadratic and cubic terms in the restoring force. The strong second harmonic component and the softening spring behaviors observed in experiment can be attributed to quadratic nonlinearity. In experiment, we apply the double-grating Doppler interferometry and achieve measurement accuracy within ten microns. We have observed the softening tuning curve and “jump phenomena”. In practical applications, the nonlinear frequency shift and jump phenomenon should be addressed especially for micro-resonators like quartz tuning forks (QTF) in various sensors and microscopies. Our experiment setup is inexpensive and easy to operate. It provides an integrated experiment for intermediate-level students and a basis for senior research projects.

\end{document}